\documentclass[twocolumn,showpacs,preprintnumbers,amsmath,amssymb]{revtex4}
\usepackage{color}
\usepackage{bm, graphicx, amsmath}
\usepackage{hyperref}

\begin{document}

\title{Scattering Quantum Random Walks on Square Grids and Randomly Generated Mazes}
\author{ Daniel Koch}
\affiliation{Air Force Research Lab, Information Directorate, Rome, New York}

\begin{abstract}

The Scattering Quantum Random Walk scheme has found success as a basis for search algorithms on highly symmetric graph structures. In this paper we examine its effectiveness at locating a specially marked vertex on square grid graphs, consisting of N$^2$ nodes. We simulate these quantum systems using classical computational methods, and find that the probability distributions that arise are very favorable for a hybrid quantum / classical algorithm. We then examine how this hybrid algorithm handles varying types of randomness in both location of the special vertex and later random obstacles placed throughout the geometry, showing that that the algorithm is resilient to both cases.

\end{abstract}


\maketitle

%
%
\section{Introduction}
%

\subsection{Quantum Random Walks}
%

Quantum Random Walks are quantum versions of classical random walks, but because of interference, their behavior can be very different \cite{davidovich,aharonov} (for reviews, see \cite{reitzner1,manoucheri}). They have proven useful in a number of algorithmic applications, one of which is searches on graphs \cite{shenvi,potocek,aaronson,lovett,childs,reitzner3,lee,feldman,hillery,hillery2,cottrell,cottrell2}. Initially the searches were for distinguished vertices, that is vertices whose behavior is different than that of the other normal vertices \cite{shenvi,potocek,aaronson,lovett,childs,reitzner3}. This has since been generalized to searches with non-uniform unmarked edges \cite{lee}, extra edges \cite{feldman,hillery2}, connections between graphs \cite{hillery}, and even a general subgraph \cite{cottrell,cottrell2}. With the recent experimental realization of discrete-time walks \cite{perets,schmitz,karski,schreiber,peruzzo,schreiber2}, it is hopeful that these and other quantum walk applications may some day soon be tested experimentally.

In more recent studies, it has been show that it is possible to use Scattering Quantum Random Walks (SQRW), a particular type of quantum random walk scheme, to create probability distributions that can aid in finding a marked vertex \cite{reitzner2,koch}. In both cases, quantum speedups were found as a result of the SQRW causing nearly all of the probability in the systems to concentrate along the path of states leading to F. This paper is an extension to the study of SQRWs, showcasing a new geometry that one can obtain a speedup on. Specifically, we extend the study of previous discrete quantum random walks on square lattices \cite{lovett2,ghosal}, now using the SQRW scheme rather than coin quantum random walks.

The predominant search algorithm studied in this paper is a hybrid quantum / classical approach, which takes inspiration from other recently successful hybrid algorithms such as the variational quantum eigensolver \cite{peruzzo2,mcclean} and quantum approximate optimization \cite{farhi,farhi2} algorithms. The goal is to use the quantum system as an aid to a classical search, whereby the results of the SQRW yield a location in the geometry that is probabilistically very near the specially marked vertex.

Lastly, we show the extent to which the SQRW scheme is still viable in scenarios with randomness in both the location of the special vertex and the geometry of the system itself. In all search algorithms the location of F is always unknown, but here we study a more difficult case where this randomness directly affects the optimal way to prepare the quantum system, unlike \cite{reitzner2,koch}. We also show how the SQRW scheme fairs under conditions where the searching geometry has unknown barriers placed throughout. These studies of randomness are motivated by more 'realistic' search problems, where classical algorithms shine due to their adaptability, but it is unclear whether or not quantum systems are viable.

\subsection{Computational Based Results}
%

Because the main focus of this paper is on the viability of a quantum search algorithm faced with new elements of randomness, there is no single graph geometry on which to conduct a full analytical study. Consequently, many of the results in this paper are generated using classical code, favoring to simulate the unitary steps done by an ideal quantum computer in order to study the resulting behaviors of the quantum system.

The major advantage to studying these quantum systems through classical simulations is that we are not limited by what kinds of geometries we can study through solely analytical means. In particular, in section 6 we simulate thousands of geometries, each of which contain up to thousands of random obstacles, making analytical results near impossible. All of the python code used to generate the results of this paper can be found at \cite{koch2}.

\subsection{Layout}
%

The layout of this paper is as follows: In section 2, we cover all of the mathematical structure for Scattering Quantum Random Walks. In section 3, we discuss the main geometry of this paper, the N$\times$N grid, and proceed to outline the searching algorithms that are best suited to utilize the kind of probability distributions generated. Section 4 is an in-depth study of the quantum systems representing these N$\times$N grids. In particular, we study how the randomness of the special vertex's location affects the quantum systems, and consequently how to adapt a hybrid search algorithm around it. In section 5, we briefly study two independent problems: an N$\times$N grid with multiple special vertices, and the 3 dimensional version of the grid, the N$\times$ N $\times$N lattice. In section 6 we return to the N$\times$N grid, placing random obstacles throughout the geometry and studying how they affect probability distributions, and consequently searching speeds. Lastly, section 7 is a concluding summary of all the results in this paper with an emphasis on geometries where SQRW based algorithms may be viable.

%
%
\section{Scatter Quantum Random Walk}
%
		
\subsection{Formalism}
%

A Scattering Quantum Random Walk is a formalism by which we can represent any graph geometry $\mathcal{G}$, consisting of vertices and connections (also referred to as nodes and edges), as a quantum system (see fig 1 below). In a classical random walk, as well as a coin-operated quantum random walk \cite{lee}, the particle is located on the vertices of the system and can travel throughout the geometry via connecting edges. For a SQRW, the edges of the graph represent the possible locations of the particle, while the vertices act as local unitary operators that propagate the walk.

Given a graph $\mathcal{G}$, consisting of $E$ edges, we can represent this geometry as a Hilbert Space $\mathcal{H}$ consisting of 2$E$ orthogonal states, two states per edge in the graph. The two states for a given edge are distinguished by their 'direction'. In particular, given two nodes A and B, there is a state $|A,B \rangle$ which represents the particle 'scattering' into node B, coming from node A, and vice versa for state $|B,A \rangle$. The direction of a state determines which local unitary operator (node) acts on it, mapping incoming states to outgoing states as follows:

\begin{equation} \label{eq:U}
U_A |j,A \rangle = -r |A,j \rangle + t \sum_{i=1,i \neq j}^{n} | A,i \rangle
\end{equation}

where $n$ is the total number of connections stemming from node A. The constants $r$ and $t$ are the reflection and transmission coefficients, given by:

\begin{eqnarray} \label{eq:rt}
t = \frac{2}{n} \nonumber \\
r =\frac{n-2}{n}
\end{eqnarray}

The unitary operation described in equation \ref{eq:U} resembles that of a scattering process, which is where this random walk scheme gets its name from. The state $|j,A \rangle$ is 'scattering' into node A, and its amplitude gets partially reflected back into the state $|A,j \rangle$ and partially transmitted into the remaining $|A,i\rangle$ outgoing states. The sum of all the local unitary operators at each vertex gives $U$, which is the unitary operator that drives the quantum system one time-step. Thus, the SQRW process is a discrete-time random walk, whereby one 'step' is defined as the action of $U$ on the entire system $\mathcal{H}$.

For all of the systems studied in this paper, the initial state of the quantum system is always an equal superposition of states, representing have no \textit{a priori} knowledge about the location of the special vertex, labeled 'F'. To drive the quantum system towards favorable probability distributions, we let the unitary operator representing the location of F, $U_F$, act with opposite phase. Specifically, this local unitary operator follows the same structure as equation \ref{eq:U}, but with +$r$ and -$t$ coefficients. This single local unitary operator is responsible for generating interference, which propagates throughout the system. After performing a desired number of unitary steps, we make a projective measurement on the system which will yield a state representing some edge.

\subsection{Constructing a Grid}
%

The main focus of this paper will be to study the geometry of a 2 dimensional (and briefly later the 3D case) N$\times$N square grid, consisting of N$^{2}$ nodes, as shown below in figure \ref{Fig:NxN}.

\begin{figure}[h]
\centering
\includegraphics[scale=0.40]{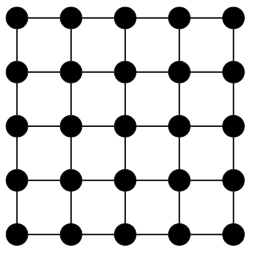}
\caption{An N$\times$N square grid, represented by a graph with nodes and edges.}
\label{Fig:NxN}
\end{figure}

Equations \ref{eq:U} and \ref{eq:rt} describe how to construct the local unitary operator for a node of $n$ connections. However, constructing these local unitary operators physically is an entirely separate challenge than simply writing them down. Specifically, as $n$ gets larger, the coefficients $r$ and $t$ become increasingly more awkward to try and implement. As an example, a node with 26 connections requires a unitary operator that can split and distribute $\frac{1}{13}$$^{th}$ of an amplitude for all 26 incoming states.

\begin{figure}[h]
\centering
\includegraphics[scale=0.38]{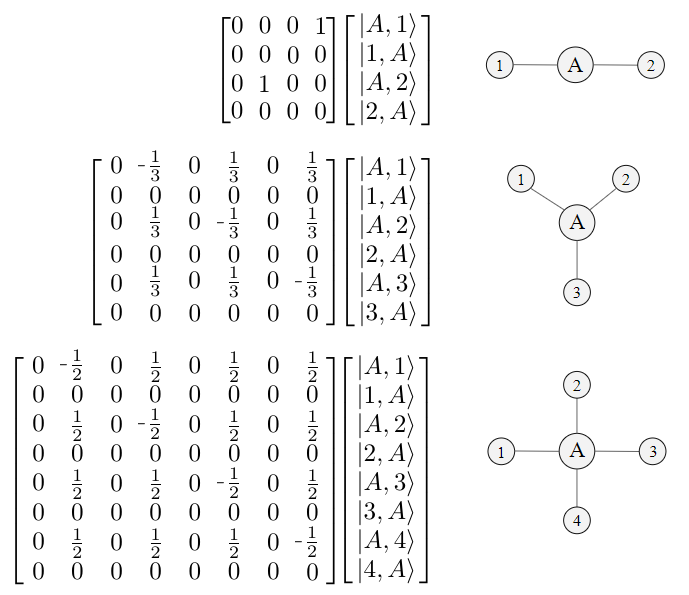}
\caption{(right) The three types of connections that are necessary to make up the N$\times$N grid graph geometry. (left) Matrices representing the local unitary operators for the centrals nodes $A$.}
\label{Fig:CM}
\end{figure}

A major motivation for the geometry studied in this paper is one that requires as few different local unitary operators as possible, and is easily scalable. The N$\times$N grids presented here only require node connections of $n$ = 2, 3, and 4, which are given by the matrices in figure \ref{Fig:CM}. Because each location is a local unitary operator, one can construct an N$\times$N grid of arbitrary size with only three core building blocks.

%
%
\section{Searching Algorithms on N $\times$ N Grids}

%

It is important to clarify exactly the type of problem we are classically representing with the use of quantum systems. In this paper, we primarily focus on the case where the exact final location is unknown, but we have complete knowledge of the graph (knowledge that we are searching on an N$\times$N grid). Later, we relax the second condition, accounting for random obstacles. In all cases, we study the effectiveness of searching for a single node, F, whose location is unknown.

\subsection{Classical Search}
%

Classically, the most effective search algorithms for a discrete graph structure, consisting of nodes and edges, are based upon either Depth-First or Breadth-First searches. The two searching techniques rely on the same underlying principles, and only differ in how they prioritize moving through the graph. Specifically, the two searches move through the graph one node at a time, keeping a running list of all nodes visited and unvisited, eventually searching through the entire graph exhaustively.

For the N$\times$N graph, our goal is to outperform the classical search in finding a specially marked vertex, which we shall refer to as F. Classically, regardless of where we start in the grid and which type of search algorithm we choose, searching for F results in the same average speed of N$^{2}$/2. This is because both algorithms simply move through the N$^{2}$ space of nodes one at a time, equivalent to a blind search through a list of entries, for which the average number of checks is half the total number of entries.

\subsection{Quantum Search}
%

In order to compare quantum vs. classical, we would first like to know how the quantum system behaves on the same geometry. In particular, following the structure outlined in section 2, we will represent an N$\times$N geometry as a Hilbert Space of orthogonal states and apply a Scattering Quantum Random Walk scheme. The hope is then to find an an ideal moment where a significant portion of the probability in the system becomes concentrated in a favorable way. 'Favorable' for example, could be that nearly all of the probability in the system concentrates on a single desired state \cite{reitzner3,lee,feldman}, or perhaps the states making up some sort of path leading to F \cite{reitzner2,koch}.

If we let only F act as a special node, meaning that it reflects and transmits with the opposite phase as all other nodes in the system, we do indeed find a moment where we achieve high probability concentrations around F. Figure \ref{Fig:PP} below shows an example of such a moment, for the case of N = 100, where F is located at the node [40,50]. In the figure, each bar represents the total probability for measuring a given node location. This total probability is the sum of the individual probabilities from the 2-4 \textit{incoming} states (one for each edge).

\begin{figure}[h]
\centering
\includegraphics[scale=0.4]{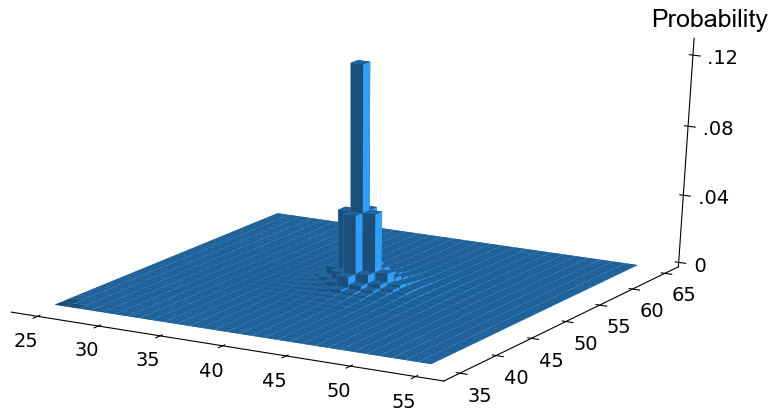}
\caption{The probability distribution P(x,y) for the case N=100, F=[40,50], after 140 unitary steps of the quantum system.}
\label{Fig:PP}
\end{figure}

Figure \ref{Fig:PP} shows the nearest 30 $\times$ 30 nodes surrounding F, rather than the full 100$\times$100. As the plot suggests, the probability does indeed continue to decrease all the way out to the edges. For this particular example, over 50\% of the total probability in the system is concentrated within a 6 node radius surrounding F, with nearly 30\% of this being accumulated on F and the four surrounding nodes alone.

Figure \ref{Fig:PP} is a good representation of many of the probability distributions that we will encounter throughout this paper. In general, for any location F in an N$\times$N grid, a SQRW will produce a moment similar to the figure, where a significant portion of the probability in the system is concentrated radially around F. But to make full use of what these quantum systems have to offer is a little tricky. As a natural first approach, we can try and imitate a Grover search, whereby the solver relies on finding F directly from a quantum measurement. Such a procedure has a simple average solving speed:

\begin{eqnarray} \label{eq:F}
P_F &\equiv& \textrm{probability of measuring node F} \nonumber \\
U_s &\equiv& \textrm{unitary steps to prepare the system} \nonumber \\
\textrm{steps}_{avg} &=& \frac{U_s}{P_F}
\end{eqnarray}

Using the data from figure \ref{Fig:PP}, $P_F \approx 0.125$ and $U_s = 140$, thus the solving speed proposed in equation \ref{eq:F} tells us that we can expect to find F in roughly 1100 steps. While this is indeed a speedup over the classical average of 5000, it is not the fastest we can do. In the nest section, we shall see that solely relying on the quantum system to find F is a waste of the system's full potential.

\subsection{ Stable Hybrid Search}
%

To motivate why a a purely quantum search is not ideal, we most note the rate at which the probability distribution depicted in figure \ref{Fig:PP} is radially decaying around F, shown below in figure \ref{Fig:N100-Pr}. For our discrete geometries, a radius is defined as the number of connections away from F. Thus, F itself is radius = 0, the four nodes surrounding F are radius = 1, and so on.

\begin{figure}[h]
\centering
\includegraphics[scale=.32]{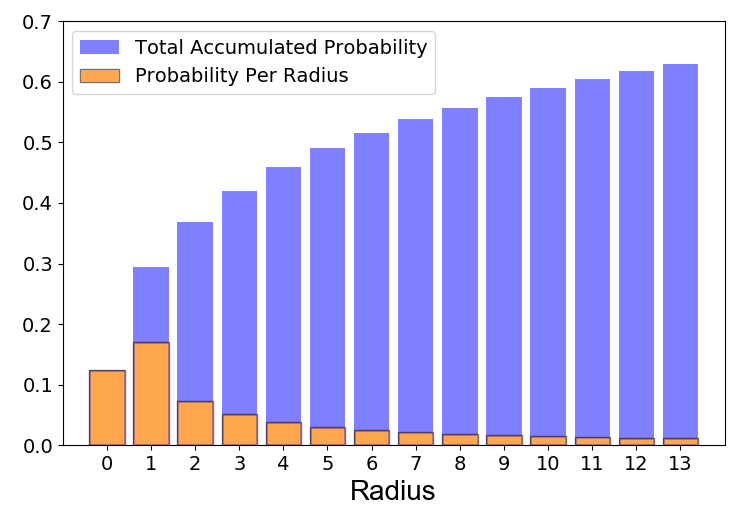}
\caption{A plot of probabilities as a function of radius, using the data from figure \ref{Fig:PP}. (blue) The total probability accumulated on all nodes up to a given radius, also referred to as P(r). (orange) The total probability accumulated on all nodes of a given radius.}
\label{Fig:N100-Pr}
\end{figure}

Figure \ref{Fig:N100-Pr} shows that the increases in probability are steadily decreasing as we move outwards from F. For a purely quantum search, the data suggests that a measurement resulting in F directly is unlikely (probability of radius = 0). However, a measurement result being 'close' to F is very favorable. But since a purely quantum search doesn't benefit from anything other than measuring F directly, measuring states close to F go to waste.

Now, suppose we make a measurement and do not find F, but can move about the grid to check nearby nodes. For this case, we can interpret the reverse of P($r$) in figure \ref{Fig:N100-Pr} as follows: if we start a search for F $\textit{from}$ the measured node, we have a P($r$) probability that we will find F within r nodes away. Therefore, if we let a classical algorithm perform a Breadth-First search starting from the location of the measurement, we can expect the algorithm to find F with the same radial probability P($r$). Figure \ref{Fig:Hybrd} below illustrates the general strategy for this hybrid algorithm approach.

\begin{figure}[h]
\centering
\includegraphics[scale=.3]{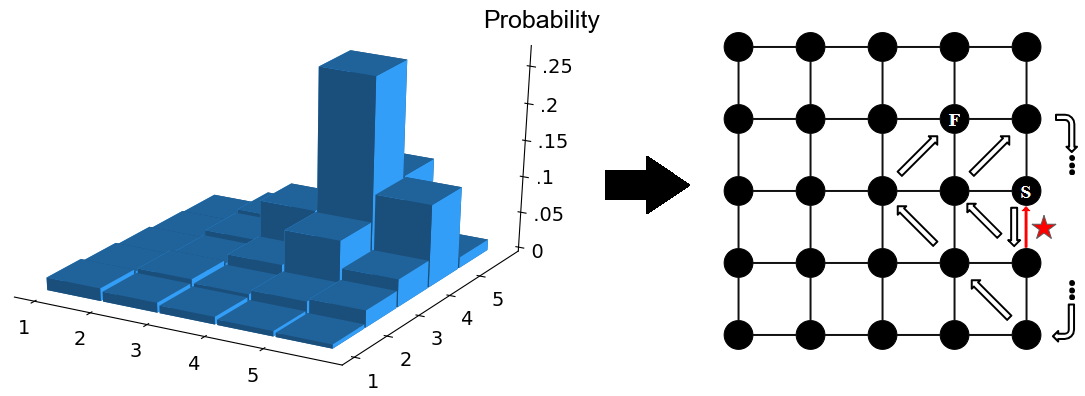}
\caption{(left) The probability distribution P(x,y) for the case N=5, F=[4,2]. (right) An example of a classical breadth-first search, following a measurement (red star).}
\label{Fig:Hybrd}
\end{figure}

The algorithm's strategy is to prepare an advantageous probability distribution P(x,y) for the quantum system (left plot), make a measurement, and then perform a classical search using the measured node (star) as our starting location. Thus, the total solving speed of this approach will be the combination of unitary + classical steps. Given the probability distribution generated from the quantum system, we can calculate the average number of steps for the resulting classical search as follows:

\begin{eqnarray}\label{eq:steps}
P(x,y) &\equiv& \textrm{probability of measuring node [x,y]} \nonumber \\
S(x,y,F) &\equiv& \textrm{classical steps to reach F, from [x,y]} \nonumber \\
\textrm{steps}_{avg} &=& \sum_{x=1}^{N} \sum_{y=1}^{N} P(x,y) * S(x,y,F)
\end{eqnarray}

where S(x,y,F) refers to the number of classical steps using a Breadth-First search.

\begin{figure}[h]
\centering
\includegraphics[scale=.32]{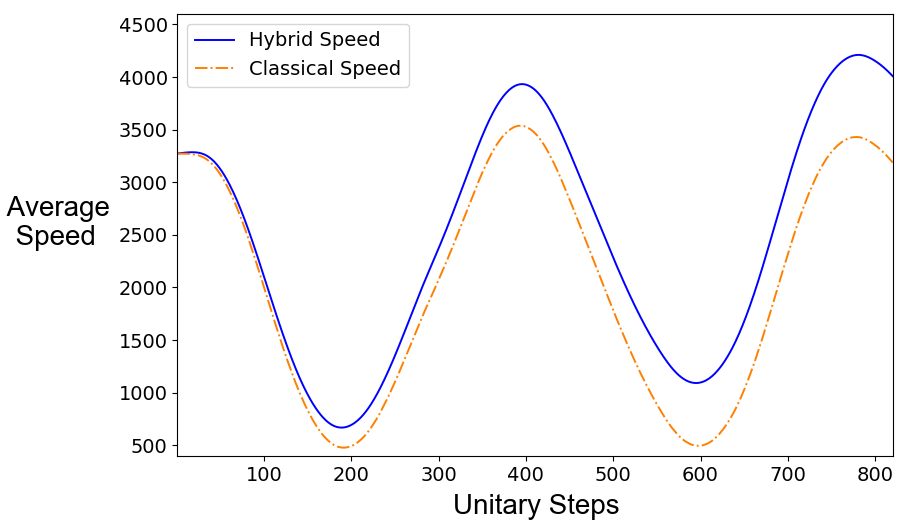}
\caption{(orange dashed) The average solving speed for a classical breadth-first search following a quantum measurement (equation \ref{eq:steps}), as a function of unitary steps on the quantum system. (blue solid) The resulting average hybrid speed (unitary + classical steps).}
\label{Fig:CSpd}
\end{figure}

Equation \ref{eq:steps} holds true for the classical case as well, where P(x,y) = $\frac{1}{N^2}$. Thus, the only difference between a purely classical and a hybrid search is P(x,y). So in order to get a speedup, we are essentially investing extra steps into the quantum system, in order to save steps classically with a favorable P(x,y). To illustrate the impact of using the SQRW scheme to produce P(x,y) distributions that result in faster classical searches, figure \ref{Fig:CSpd} shows equation \ref{eq:steps} as a function of unitary steps of the system, for the case N=100, F=[40,50].

Figure \ref{Fig:CSpd} shows that by using the quantum system to prepare a desirable P(x,y), one can significantly reduce the average number of steps expected from the resulting classical search. For the example in the figure, investing $\sim$ 200 unitary steps into the quantum system results in saving over 2500 classically on average. Together, the fastest hybrid speed is a combination of 190 unitary steps + an average of 480 classical steps, for a combined hybrid average speed of 670. This is roughly 40\% faster than the purely quantum search speed, and over 7 times faster than the classical.

Equation \ref{eq:steps} is a summation over all the nodes in the system, meaning that it represents the case where we let the classical algorithm search exhaustively until it finds F. But as shown in figure \ref{Fig:N100-Pr}, the majority of the weight in this average comes from the nodes clustered around F. Searching further and further away from the starting node has diminishing returns. Thus, as we shall see in the next section, at a certain point it actually becomes advantageous to stop the classical search and start over again with a new quantum system and measurement.

However, the hybrid algorithm in this section has a noteworthy property, despite not being the fastest algorithm studied in this paper. Namely, a speedup is achievable with only one preparation of the quantum system. If one considers the technological state of current quantum computers (including errors and loss), it may be unrealistic to favor more 'quantum demanding' approaches that require quantum systems we can repeatedly create, perform unitary steps on, and measure. In addition, because this approach uses an exhaustive classical search, the hybrid search has a finite maximum of N$^2$ + $U_s$ steps. For this reason, we will shall refer to this style of hybrid search, with an exhaustive classical search component, as a 'stable hybrid search.'
\subsection{Optimal Hybrid Search}
%

The stable hybrid and the purely quantum approaches studied thus far essentially represent the same technique, differing only in how much one searches radially around the location of the measurement. The stable search represents the maximum case, while the quantum search represents the minimum (radius of 0).

In this final algorithm, we will try and strike a balance between these two approaches. Specifically, after making a quantum measurement, we proceed with a classical search for F within a finite radius. If we do not find F within this radius, we prepare a new quantum system and repeat the process. The average speed at which we can expect to find F, for a given searching radius $r$, is as follows.

Let us define the following quantities:

\begin{eqnarray}\label{eq:terms}
r_{max} &\equiv& \textrm{maximum radius for the classical search} \nonumber \\
\textrm{S}_{max} &\equiv& \textrm{average maximum classical steps per search} \nonumber \\
\textrm{S}_{F} &\equiv& \textrm{average steps to F in a successful r}_{max} \textrm{ search} \nonumber \\
\textrm{U}_s &\equiv& \textrm{unitary steps to prepare the system} \nonumber \\
&&\textrm{P}_{success} = P(r_{max}) \\
&&\textrm{P}_{fail} = 1 - \textrm{P}_{success}
\end{eqnarray}

where P$(r)$ is the total probability accumulated within $r$ nodes of F (see figure \ref{Fig:N100-Pr}).

In order to calculate the average hybrid speed for a given radius, we must factor in the possibility of failures, which result in 'wasted' steps. When the classical search does not yield F, the search suffers an average of S$_{max}$ + U$_s$ wasted steps. For the trial that succeeds in finding F, we have S$_{F}$ + U$_s$ total steps, where S$_{F}$ is equivalent to equation \ref{eq:steps}, but only consisting of nodes within the searching radius:

\begin{equation}\label{eq:Sc}
\textrm{S}_{F}= \sum_{x} \sum_{y} \frac{P(x,y) \cdot S(x,y,F)}{\textrm{P}_{success}} \hspace{.4cm} [x,y] \subset r_{max}
\end{equation}

All together, the average hybrid speed has the following form:

\begin{equation}\label{eq:Hspd}
\textrm{Speed}_{avg} = \textrm{P}_{success} \sum_{k=1}^{\infty}( \textrm{P}_{fail} )^{k-1} \cdot [ \textrm{Steps}(k) ]
\end{equation}

where Steps($k$) is the number of total steps after exactly $k$ quantum measurements. Specifically, if we find F on the $k^{th}$ attempt, following $k-1$ failures, then the total number of steps will be:

\begin{equation}\label{eq:S(k)}
\textrm{Steps}(k) = (\textrm{U}_s + \textrm{S}_{max})\cdot (k-1) + \textrm{U}_s + \textrm{S}_{F}
\end{equation}

Equation \ref{eq:Hspd} has the following mathematical form:

\begin{equation}\label{eq:sum}
\sum_{k=1}^{\infty}(p)^{k-1} \cdot [ A(k-1) + B ]
\end{equation}

which converges to:

\begin{equation}\label{eq:sum2}
\frac{Ap - Bp + B}{(1-p)^2} \hspace{.6cm} |p| < 1
\end{equation}

where A = $\textrm{U}_s + \textrm{S}_{max}$, B = $\textrm{U}_s + \textrm{S}_{F}$ and $p = \textrm{P}_{fail}$. Plugging in these values, we get the following equation for the average hybrid search speed, for a given radius:

\begin{equation}\label{eq:sum3}
\frac{\textrm{P}_{fail} \cdot (\textrm{S}_{max} - \textrm{S}_F) + [U_s + \textrm{S}_F]}{(\textrm{P}_{success})^2}
\end{equation}

\begin{figure}[h]
\centering
\includegraphics[scale=.32]{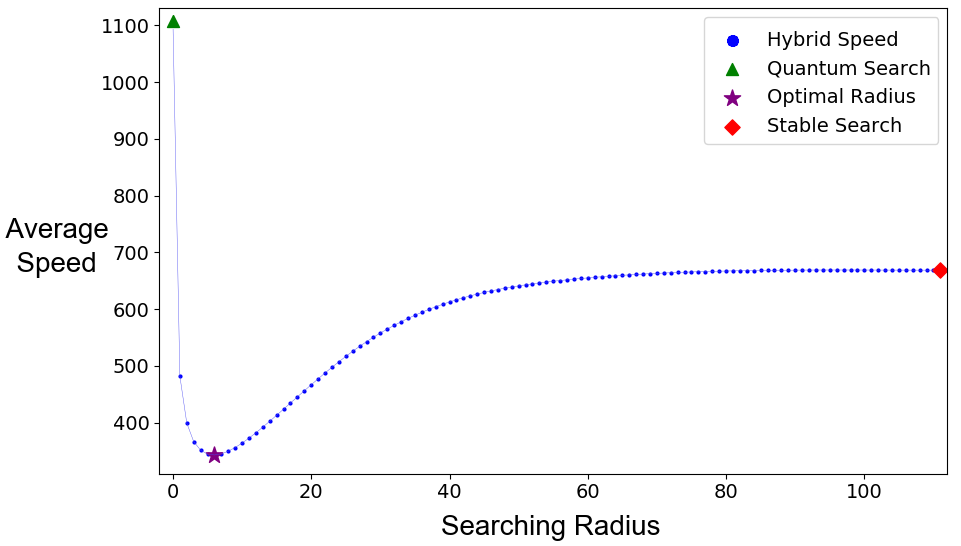}
\caption{(circles) The fastest hybrid speeds as a function of searching radii $r$. (triangle) The purely quantum search. (diamond) The stable hybrid search. (star) The optimal searching radius.}
\label{Fig:HSvR}
\end{figure}

Equation \ref{eq:sum3} shows the balance of choosing an optimal searching radius. For a small $r$, the step counts from S$_{max}$ and S$_{F}$ are minimized, but the resulting small P$_{success}$ term in the denominator becomes problematic. Conversely, going to larger searching radii has diminishing returns on P$_{success}$ (see figure \ref{Fig:N100-Pr}), and the S$_{max}$ and S$_{F}$ terms drive the step count up. Thus, the radius corresponding to the fastest hybrid speed will be the one that minimizes equation \ref{eq:sum3}. This is shown in figure \ref{Fig:HSvR}, for the case N=100, F=[40,50]. Also note that for P$_{fail}=0$, equation \ref{eq:sum3} simply becomes the stable hybrid approach studied earlier.

As expected, the fastest hybrid searching speed comes from some intermediate value of r between r=0 (the purely quantum search) and the maximum (the stable hybrid search). For this particular case, $r$ = 6 turns out to be the best case, incorporating a total probability of 61\%, requiring only 167 unitary steps, and achieving an average solving speed of 343. This is almost twice as fast as the stable search, and nearly four times faster than the quantum search. In general, the fastest hybrid speed is always some intermediate radius, dependent on both the grid size N and the location of F.

\subsection{Grid Size Trends}
%

\begin{figure}[h]
\centering
\includegraphics[scale=.34]{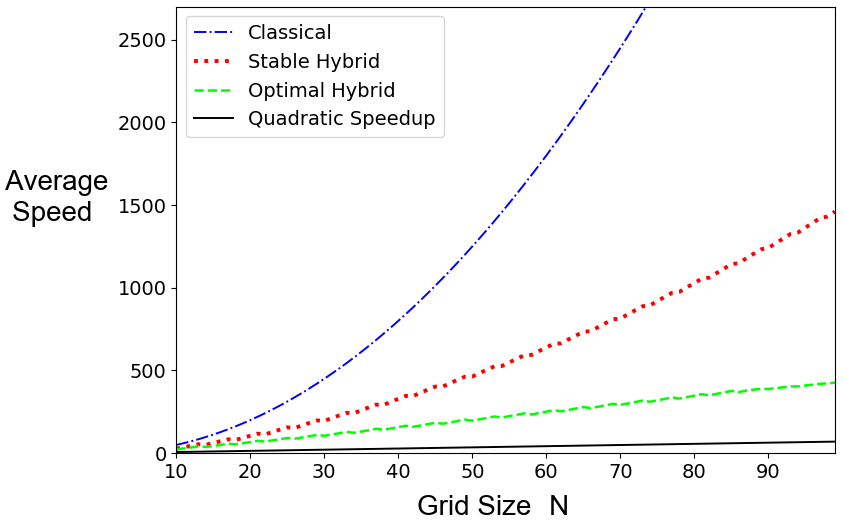}
\caption{Fastest average searching speeds as a function of grid size N. (solid line) Plotted for reference is a quadratic speedup over the classical search: $\sqrt{N^2 / 2}$}
\label{Fig:Ntrends}
\end{figure}

The example used in this section for N=100 was for a single F, and one close to the center of the grid to showcase good results. In the next section, we shall see that solving speeds are heavily dependent on F's location. But before moving on, it is worth noting how these solving speeds depend on the size of the grid. To show this, figure \ref{Fig:Ntrends} shows the fastest stable and optimal hybrid speeds as a function of N, for F located at the node closest to [N/4, N/4].

The trends in figure \ref{Fig:Ntrends} are pretty straightforward, showing that these hybrid searches provide some degree of power-like speedup, certainly slower than a quadratic though. The choice for F=[N/4, N/4] is motivated by a result shown in the next section. Specifically, it represents a good 'average' location that results in speeds close to what one might expect when doing a blind search.

%
%
\section{Blind Hybrid Search}
%

For an exactly known system, like the examples in section 3, one can always opt to measure at the most advantageous time. For example, if you \textit{know} that you have a 100$\times$100 grid and F is located at [40,50], then all of the claims about speeds from the previous section hold true. However, for systems with only partial information, such as an N$\times$N grid with an unknown F, we cannot prepare the system for a single optimal moment. Instead, we must take into consideration all of the possible F locations and how they impact our decision of when to measure the quantum system.

\subsection{Dependence on F}
%

When designing an algorithm around quantum systems, which naturally have their own degree of randomness, additional restrictions on information can be problematic. For example, a Grover search has a very well known peak for N list entries, but how do you prepare your system if the number of entries is unknown?

\begin{figure}[h]
\centering
\includegraphics[scale=0.3]{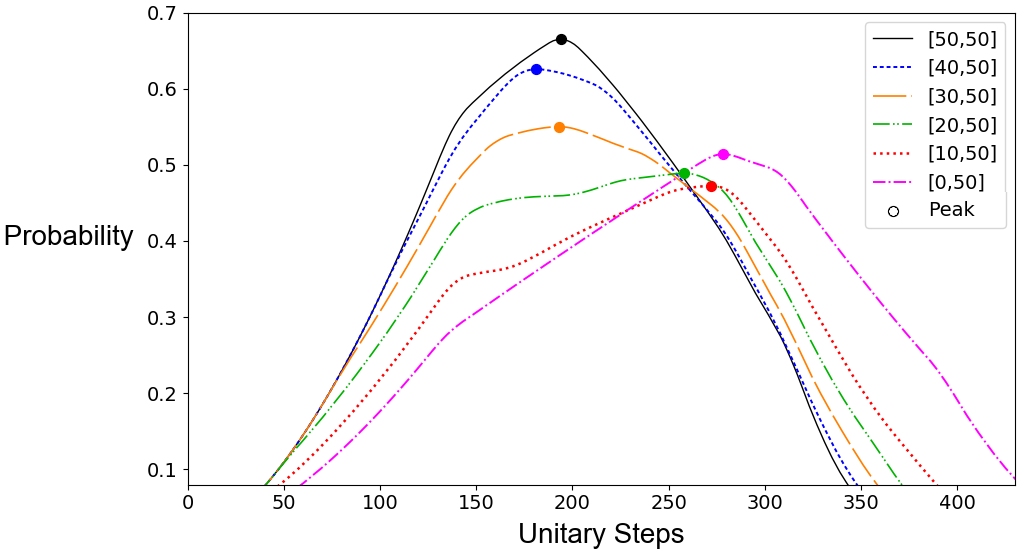}
\caption{(plots) The total probability accumulated within a 6 node radius of F, for various F locations, as a function of unitary steps. (circles) The moment of the highest total probability.}
\label{Fig:6F}
\end{figure}

\begin{figure}[h]
\centering
\includegraphics[scale=0.28]{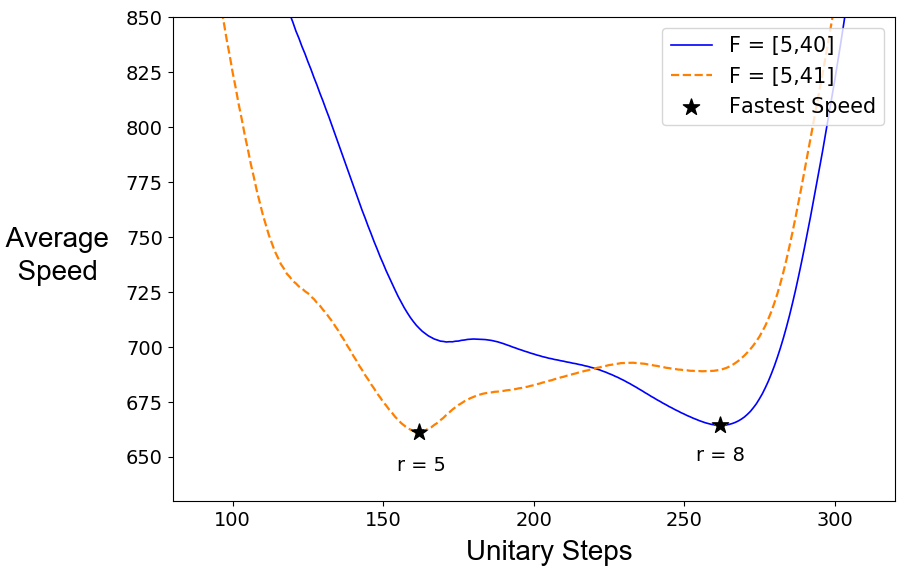}
\caption{Plots of optimal hybrid speeds as a function of unitary steps, for F locations [5,40] (solid) and [5,41] (dashed). Next to each plot is the optimal searching radius, as well as the corresponding fastest speeds (stars)}
\label{Fig:Comp}
\end{figure}

For these N$\times$N grids, we would ideally like the system to behave exactly the same way regardless of where F is located. Then, we could always prepare any N$\times$N system the same way, and be guaranteed optimal results. Unfortunately, this is not the case. To illustrate this point, figures \ref{Fig:6F} and \ref{Fig:Comp} show that different locations of F have different probability trends and peak moments.

Figures \ref{Fig:6F} and \ref{Fig:Comp} show that depending on F's location in the grid, the resulting P(x,y) probability distributions can vary quite drastically in shape and peak. As a result, F's location directly affects the ideal number of unitary steps to prepare the quantum system as well as the optimal searching radius. Consider the example shown in figure \ref{Fig:Comp}, and the difference between two neighboring nodes.

Since we cannot look for a single moment to optimize for, we must instead factor in all of the possible locations for F, and pick a moment that gives us the best average. That is to say, supposing F could be anywhere in the grid, what is the number of unitary steps that we should prepare our quantum system for, in order to give ourselves the best P(x,y) for finding a node closest to F. In addition, if we wish to use an optimal hybrid search algorithm, we must also factor in how F's location affects the best search radius $r$.

\subsection{Optimal Measurement for single F location}
%

In order to see if there is a single combination of unitary steps and optimal search radius that we can use for any given N$\times$N, we need to study all possible F locations. In this section and the next, we will study the case for N=100, and use it to judge the viability for a blind search.

Although there are N$^2$ nodes in a system, the number of 'unique' nodes is roughly 1/8th of this number. Specifically, if we characterize a node by its distances to the four walls of the grid, there are up to 8 nodes with the same characterization (nodes on the diagonals have 4 symmetric locations, while all other nodes have 8).

For example, N=100, the nodes located at [0,1], [1,0], [99,0], [100,1], etc. are all equidistant from the four walls. For the quantum system, these locations will all produce the exact same P(x,y) distributions, as they are all geometrically symmetric to each other. Thus, for a complete analysis of an N$\times$N grid, we need only study all of the unique nodes, which can be thought of as a single octant of the total grid.

For an N$\times$N system, the number of unique nodes is:

\begin{eqnarray}\label{eq:Unq}
\frac{1}{8} N(N+2) \hspace{1cm} &&N=even \nonumber \\
\frac{1}{8} (N+1)(N+3) \hspace{1cm} &&N=odd
\end{eqnarray}

\begin{figure}[h]
\centering
\includegraphics[scale=.30]{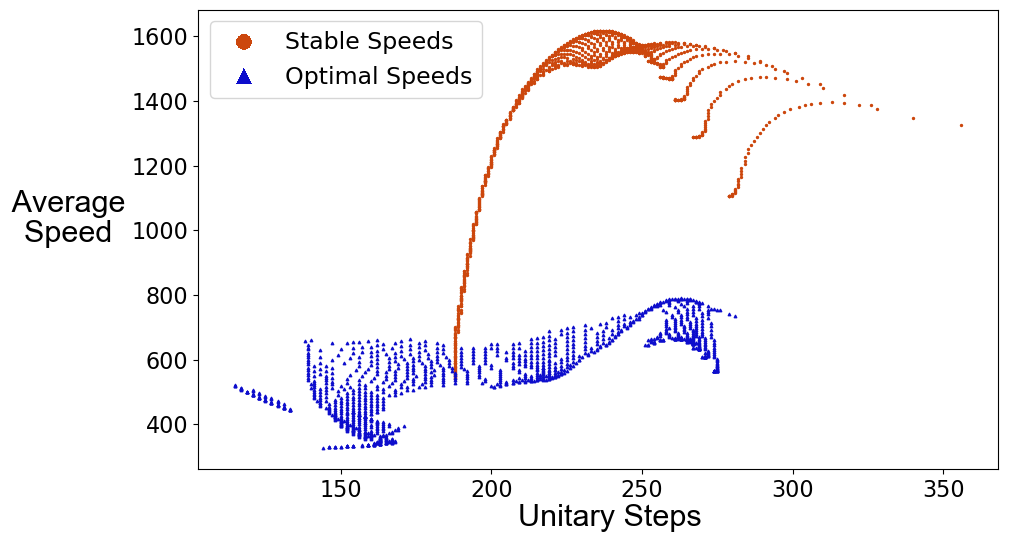}
\caption{A scatter plot of fastest hybrid speeds vs unitary steps, for the 1275 unique F locations for the case N=100. (circles) Scatter plot of the fastest stable hybrid speeds. (triangles) Scatter plot for the fastest optimal hybrid speeds.}
\label{Fig:AllF}
\end{figure}

For the case of N=100, we have 1275 unique nodes. Using these 1275 locations, we let a classical computer calculate the fastest stable and optimal hybrid speeds, which are plotted in figure \ref{Fig:AllF}.

We find that the stable hybrid search speeds (circles) vary significantly in both fastest speeds and corresponding unitary steps. The data points seen in 'arch-like' shapes in the upper right corner of the figure correspond to nodes closer to the boundaries, while nodes closer to the center of the grid can be seen by the tight vertical cluster of data points with unitary step counts ranging from 180-220.

\begin{figure}[h]
\centering
\includegraphics[scale=.30]{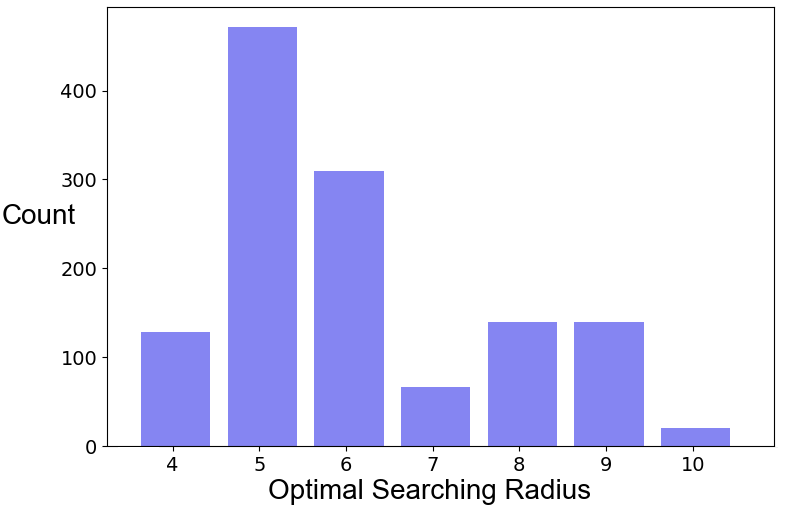}
\caption{A histogram of the optimal searching radius $r$ corresponding to the fastest hybrid search speeds plotted in figure \ref{Fig:AllF}. For N=100, the mode is $r$=5, while the mean is closest to $r$=6.}
\label{Fig:Allr}
\end{figure}

By comparison, the optimal hybrid search speeds (triangles) appear to have a much more consistent spread in speeds, but an equally large variance in unitary step values. However, this is because figure \ref{Fig:AllF} shows the fastest hybrid search speeds corresponding to the optimal search radius for each F, which vary as well. Figure \ref{Fig:Allr} shows a count of optimal search radii.

Even though the optimal hybrid search speeds are more consistent, figure \ref{Fig:Allr} reveals that the corresponding optimal search radii are not. If we recall the plots from figure \ref{Fig:Comp}, where the location of F caused two neighboring nodes to have significantly different peak moments and optimal searching radii, we can see this effect over the whole grid. Specifically, nodes closer to the center of the grid tend to favor searching radii of 4-6, while nodes closer to the boundaries favor 8-9. This split is partly due to the fact that F locations closer to the boundaries need larger searching radii in order to move through the same number of nodes as more centralized F locations.

\subsection{Blind Optimal Measurement}
%

\begin{figure}[h]
\centering
\includegraphics[scale=.32]{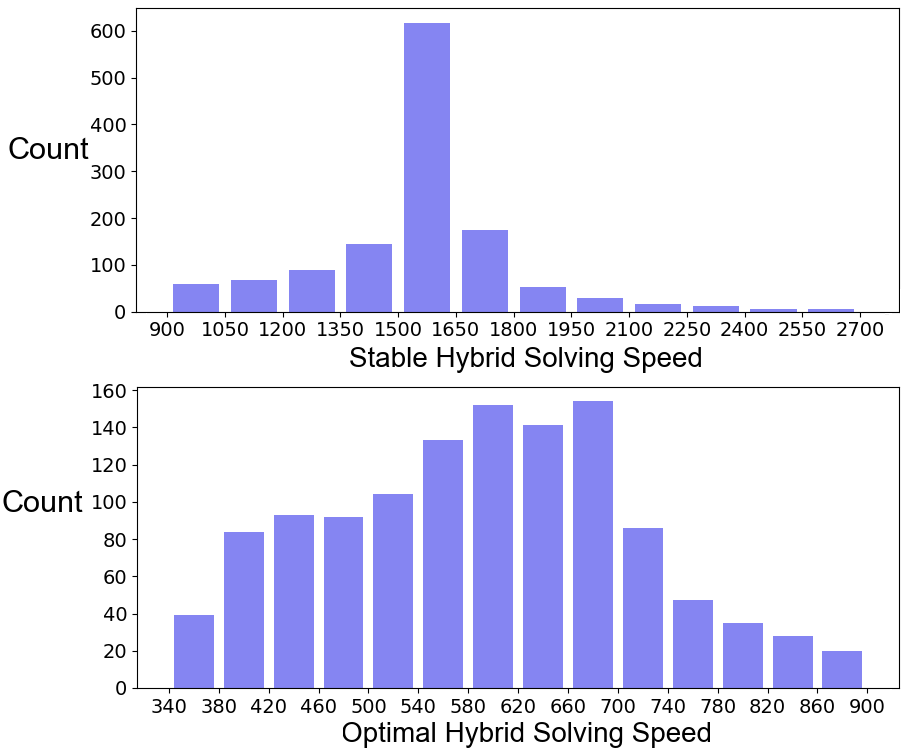}
\caption{(top) Histogram of fastest stable hybrid speeds. (bottom) Histogram of fastest optimal hybrid speeds.}
\label{Fig:Blind}
\end{figure}

Based on the results from figures \ref{Fig:AllF} and \ref{Fig:Allr}, we would now like to ask whether or not a viable hybrid search can be conducted from a single unitary step count and search radius. If we average the data points from these figures properly (using the 1275 unique nodes to represent the total 10,000), we find that for the stable hybrid search we need 232 unitary steps, and for the optimal hybrid search we want 199 unitary steps with a search radius of 6. Using these values, we again let a computer calculate the average solving speeds, following equations \ref{eq:terms} - \ref{eq:sum3}. The results are shown in figure \ref{Fig:Blind}.

Beginning with the top plot in Figure \ref{Fig:Blind}, the stable hybrid approach, we see a much larger range of speeds, but generally much more clustered around a single value. Conversely, the optimal hybrid approach leads to a much more even distribution over a smaller range. If we average the values in these plots, we get our final answers: that the 'blind stable hybrid' search gives an average search speed of approximately 1560 total steps, while the 'blind optimal hybrid' search, of search radius 6, has an average speed of approximately 475. Comparing to the classical average of 100$^2$/2, we find that both approaches are a speedup, with the optimal cases being several times faster.

\begin{figure}[h]
\centering
\includegraphics[scale=.32]{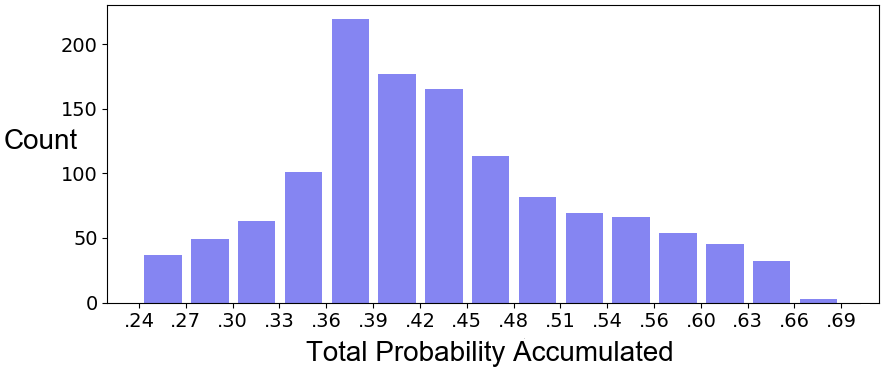}
\caption{Histogram of the total probability accumulated within a 6 node radius of each F, for all the unique nodes in the blind optimal search (figure \ref{Fig:Blind}).}
\label{Fig:Pcts}
\end{figure}

To highlight why the optimal searching radius of 6 is so resilient to the randomness of F's location, we need only look at the P($r=6$) probabilities. Figure \ref{Fig:Pcts} shows a count of how much of the probability in the system is accumulated for the blind optimal searches from figure \ref{Fig:Blind}

As shown by the distribution, no matter where F lies in the system, the blind optimal hybrid search will sweep through enough nodes to have good probability of finding F. Figure \ref{Fig:Pcts} shows that the average chance of finding F within any given trial is about 40\% - 45\%, which is certainly viable enough to use as a basis for the quantum component of a hybrid algorithm.

This concludes our study on the case of N=100. But as a final remark, it is worth noting that N=100 is nothing special. Based on figure \ref{Fig:Ntrends}, there is no reason to suspect that the techniques and resulting success studied in this section would not scale to larger sizes, and even smaller sizes down to a certain threshold. N=100 is large enough to see the potential for a quantum speedup, and based on studies of different geometries like \cite{reitzner2,koch}, one should expect similar power-like scaling results for larger N$\times$N grids.

%
%
\section{Aside: Geometries of Interest}
%

In this section we would like to present some results based on problems that are closely related to the N$\times$N grid searches studied up to this point. The following subsections are stand alone results, but contribute to the overall discussion in the final conclusion section of this paper.

\subsection{Two Specially Marked Nodes}
%

Let us consider the problem of pathfinding, say between two nodes A and B. For these discrete grids, where each node can have up to four connections, finding the shortest path from point A to B follows the rules of taxicab geometry. Classically, if there are no obstacles within the grid, there are many equally short paths from A to B, with the exception being when they share the same x or y coordinate.

For the quantum system, having a starting and final node means we now have two specially marked vertices that transmit and reflect with opposite phase. We have seen that letting a single node be special results in radially concentrated probability distributions, so let us now see what happens with two. Figure \ref{Fig:SF} below shows the case for N=20, S=[6,20], F=[12,20], displaying a plot of P(x,y) for the entire grid.

\begin{figure}[h]
\centering
\includegraphics[scale=.36]{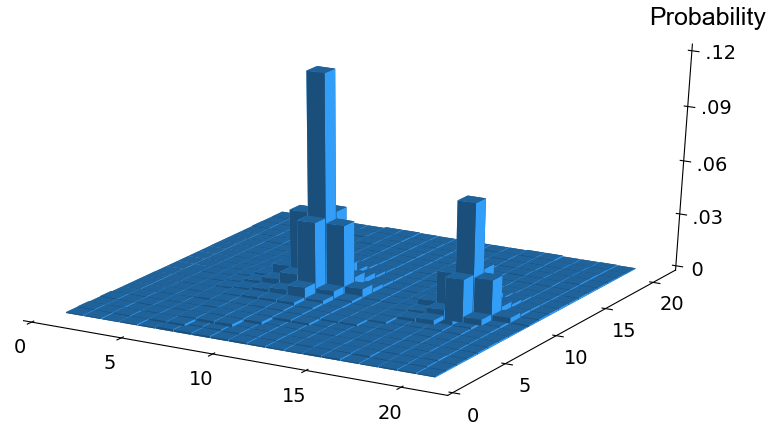}
\caption{The probability distribution P(x,y) for the case N=20, with two specially marked vertices at locations [6,20] and [12,20]. The probability in the system concentrates around the two locations, similar to if either location were the only special vertex in the system.}
\label{Fig:SF}
\end{figure}

Unfortunately, the probability in the system concentrates around S and F, but shows no sign of the path between the nodes. While this result is interesting, implying that the SQRW scheme may be useful in locating multiple special nodes, it does not help us in terms of a path-finding search. Nevertheless, we present this result to illustrate what other possible types of problems may be solvable.

\subsection{ N$\times$N$\times$N Lattice }
%

Besides just 2D graphs, it is worthwhile to study how the SQRW scheme performs on higher dimensional geometries. As the natural choice from a grid, we will present some results generated on cubic lattices, as shown below in figure \ref{Fig:cube}. The quantum systems representing these cubes follow all of the same structure as outlined in section 2, only now we have up to six connections on a single node.

\begin{figure}[h]
\centering
\includegraphics[scale=.30]{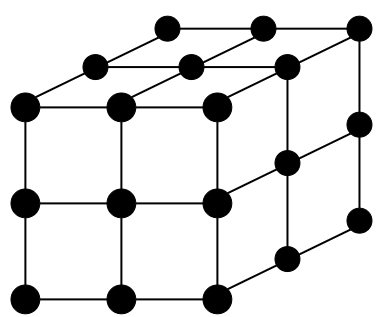}
\caption{An N$\times$N$\times$N cubic lattice, represented by a graph with nodes and edges.}
\label{Fig:cube}
\end{figure}

Our problem is still to find a specially marked vertex F whose location is unknown. Classically, now we have N$^3$ nodes to search through, resulting in an average search speed of N$^3$/2. Here, we will not go through the same rigorous study of these geometries as with the 2D grids, but rather just a few quick observations and trends.

Similar to figure \ref{Fig:N100-Pr}, the effect of the SQRW on N$\times$N$\times$N lattices results in radially decaying probability distributions. Because there are more nodes per radius for the 3D case, the total accumulated probabilities are naturally higher. Thus, to compare the grid and cube geometries properly, we will go directly to their hybrid searching speeds.

\begin{figure}[h]
\centering
\includegraphics[scale=.5]{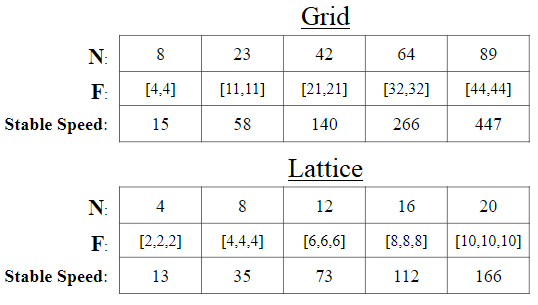}
\caption{A comparison of grid and lattice geometries with similar total nodes. The results indicate that for two geometries containing the same number of total nodes, assembling the nodes in a cubic lattice yields better searching results than a square grid.}
\label{Tbl:GvC2}
\end{figure}

In order to isolate the how the 3D geometry impacts searching speeds, we want to compare geometries with roughly the same number of total nodes in the system. This way, both geometries will have comparable classical searching speeds. Figure \ref{Tbl:GvC2} shows two tables with various lattice and corresponding closest grid sizes, along with their resulting stable hybrid speeds.

Since the systems in figure \ref{Tbl:GvC2} have nearly the same number of total nodes, we can attribute the faster solving speeds to the geometry of the lattice. Unsurprisingly, with the same number of nodes, we find that the system with more symmetry results in a higher concentration of probability around F, which leads to faster hybrid speeds. If we extrapolate the results from the blind search in section 4, to 3D lattices, there is good evidence to suggest similar success.

%
%
\section{Walls and Mazes}
%

We now return to the study of N$\times$N grids here. Based on the results presented in sections 3 and 4, grids are prime candidates for using a hybrid search algorithm. In this final section, we will test the viability of SQRWs as a means for searching on geometries with random obstacles.

The goal is to study maze-like geometries on N$\times$N grids by adding walls randomly throughout the space, and seeing how it affects probability distributions. We will study up to the case with the maximum number of walls placeable in an N$\times$N grid, where there is exactly one path from any node [x,y] to any other node [x',y']: the perfect square maze. We have already seen that letting two nodes act as special vertices does not reveal any sort of path, thus we will continue to only have a single F in the system.

\subsection{N$\times$N Grid with Walls}
%

Let us first define what we mean by adding walls to the geometry of the system. A wall will represent a boundary between two neighboring nodes [x,y] and [x',y']. Figure \ref{Fig:GwW} below shows an example of a 4$\times$4 grid with two walls touching the node [2,3], removing it's connection from nodes [1,3] and [2,4]. As a result, the states representing those edges are no longer a part of the quantum system, and correspondingly the classical search algorithm can no longer directly travel between those nodes.

\begin{figure}[h]
\centering
\includegraphics[scale=.4]{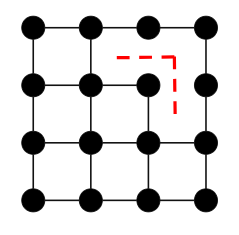}
\caption{An N$\times$N square grid with walls separating neighboring sites.}
\label{Fig:GwW}
\end{figure}

We will first test to see how these walls affect the probability distributions P(x,y), and correspondingly our hybrid search speeds. Ultimately, we want to know how resilient these quantum systems are to random obstacles placed throughout the grid. These obstacles could represent physical boundaries in a geometry we are searching on, or perhaps unintended boundaries created in a non-perfect quantum system, where errors in qubits cause local unitary operators to behave improperly.

We will study the case of N=40, F=[10,15], adding in walls to the geometry in regular increments. For each increment of walls added into the system, we want to study as many randomly generated permutations as we can, and see on average how they affect solving speeds. The results generated in this section reflect 500 randomized geometries per wall increment. The only condition we have for placing walls is that they never section off any nodes from the rest of the system. This way, there always exists at least one path connecting any two nodes, and no probability in the system is ever 'trapped'.

Figures \ref{Fig:Walls} and \ref{Fig:H-Walls} below show scatter plots of the fastest stable and optimal hybrid speeds vs. unitary steps, for select amounts of walls in the geometry. Each data point comes from a different randomly generated maze geometry, used for both hybrid searches.

\begin{figure}[h]
\centering
\includegraphics[scale=.32]{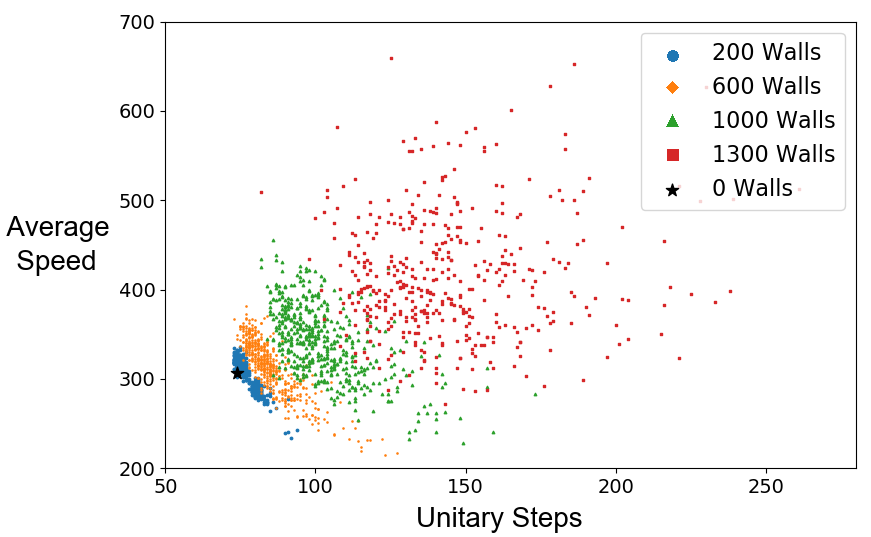}
\caption{A scatter plot of the fastest stable hybrid speeds vs unitary steps, for randomly generated geometries containing walls (see fig \ref{Fig:GwW}), for the case N=40, F=[10,15].}
\label{Fig:Walls}
\end{figure}

\begin{figure}[h]
\centering
\includegraphics[scale=.32]{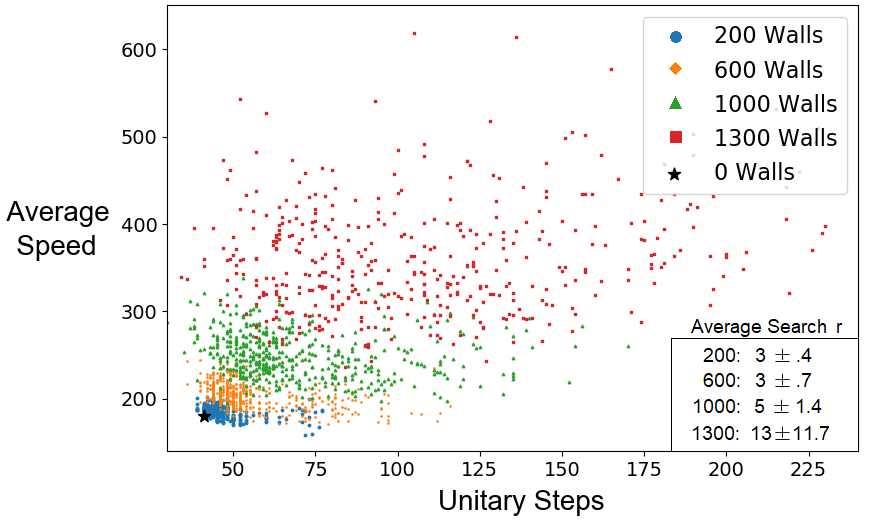}
\caption{ A scatter plot of the fastest optimal hybrid speeds vs unitary steps, for the same randomly generated geometries as studied in fig \ref{Fig:Walls}. While typically faster than the stable searches, the data points tend to have higher degrees of variance in unitary steps. }
\label{Fig:H-Walls}
\end{figure}

As shown by the distinguishable clusters of points, adding random walls into the system does indeed cause variations in fastest solving speeds and unitary steps. In general, we find that for both algorithms, adding walls into the system seems to almost always drives the optimal number of unitary steps up, as compared to the case with zero walls (black star). However, sometimes the presence of walls in the system actually results in \textit{faster} average searching speeds (shown in figure \ref{Fig:Walls} by points lower than the black star).

For a reference, the maximum number of walls that can be placed into an N=40 grid is 1521 (becoming a perfect square maze). By averaging the data points collected from all the randomized geometries, figure \ref{Fig:Avgs} below shows a comparison of the stable and optimal search algorithms as a function of walls in the system. For lower wall counts, the optimal search is still significantly faster, and has much less variation in speeds. However, by 1400 walls in the system, we find that the optimal hybrid offers almost no advantage.

\begin{figure}[h]
\centering
\includegraphics[scale=.32]{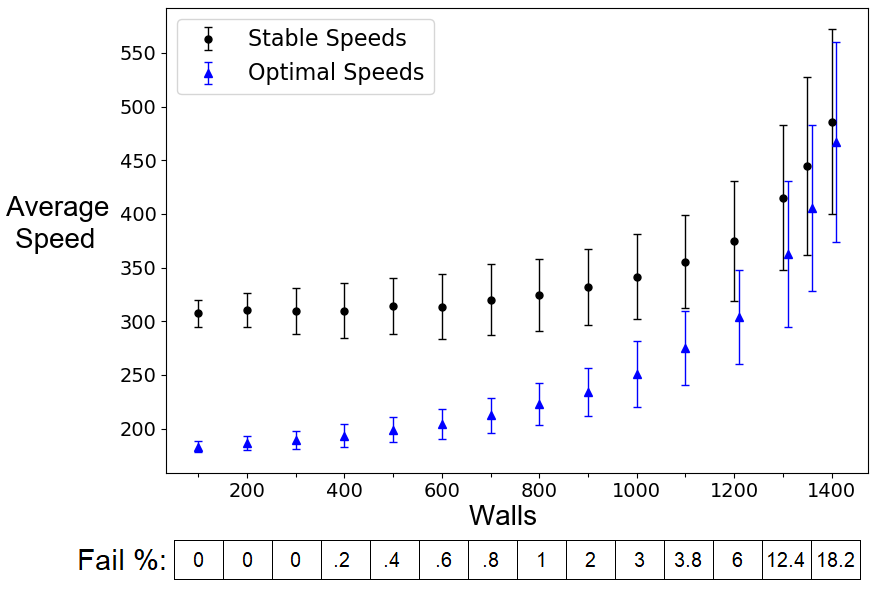}
\caption{(data points) A plot of average solving speeds as a function of walls in the grid geometry, for N=40, F=[10,15], including the data from figures \ref{Fig:Walls} and \ref{Fig:H-Walls}. (error bars) The error bars on each data point represent one standard deviation. (below) A table showing the failure rate for each wall count, whereby neither the stable or optimal hybrid searches obtained a speedup. Note: only non-failure data points were used for the calculations of averages and standard deviations in the above plot.}
\label{Fig:Avgs}
\end{figure}

Although the optimal hybrid search achieves similar success over the stable search for low wall counts, figure \ref{Fig:Avgs} show that its viability starts to rapidly evaporate as the wall counts approach the maximum. Note at the bottom right corner of figure \ref{Fig:H-Walls}, that the variation in optimal searching radii is becoming so large that is is causing the standard deviation to be nearly the same as the average. This variation in optimal searching radii kills any chance of success for a blind optimal hybrid algorithm.

Below the plot in figure \ref{Fig:Avgs} is a table showing 'Failure $\%$'. This is the percentage of geometries that were studied, where neither the stable or optimal hybrid searches could produce a speedup. The numbers in each box correspond to wall counts in increments of 100, meant to line up with the x-axis. For the averages and standard deviations plotted above, data points that were not a speedup were removed from the dataset. Thus, the plot in figure \ref{Fig:Avgs} is based on $\textit{successful}$ searches, meant to showcase what a solver might expect for the successful attempt.

As it turns out, 1400 walls is very near the limit to where an optimal searching radius is still possible, although practically unviable. Beyond 1400, it is accurate to say that the optimal hybrid search becomes the stable hybrid search, where the optimal searching radius is simply the maximum. The optimal optimal hybrid approach becomes essentially unadvantageous well before this limit however, around the 1100-1200 wall count for N=40, which corresponds to around 75$\%$ of the maximum.

While the data from figure \ref{Fig:Avgs} seems to rule out any hope at a full maze solving algorithm (as we shall see in the next section), it is still noteworthy as to how resilient the SQRW approach is at lower wall counts. In fact, even up to around 50$\%$ of the maximum, 600-800 walls, both the stable and optimal hybrid approaches are still very viable. Thus, figure \ref{Fig:Avgs} is actually quite reassuring that there is still potential for the SQRW technique on geometries with randomness, which we will comment on in the conclusion section.

\subsection{Perfect Square Mazes}
%

To conclude our study on grid geometries with randomly generated walls, we will study the limit to these geometries: perfect square mazes. A perfect square maze is categorized such that there exists exactly one path between any two nodes in the system. An example of such a maze geometry is depicted below in figure \ref{Fig:PSM}

\begin{figure}[h]
\centering
\includegraphics[scale=.28]{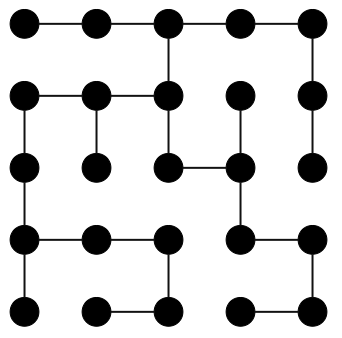}
\caption{A graph representing of a perfect square maze as nodes and edges.}
\label{Fig:PSM}
\end{figure}

As noted before, for N=40, the optimal hybrid approach loses all viability around 1400 walls, which corresponds to around 92$\%$ of the maximum. For completeness, figure \ref{Tbl:PSM} below shows the failure rate of the stable hybrid approach as the number of walls in the system approaches the limit.

\begin{figure}[h]
\centering
\includegraphics[scale=.6]{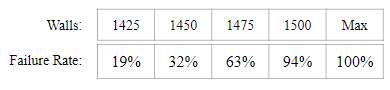}
\caption{The percentage of geometries studied where a speedup over the classical average N$^2$/2 was unobtainable.}
\label{Tbl:PSM}
\end{figure}

\begin{figure}[h]
\centering
\includegraphics[scale=.33]{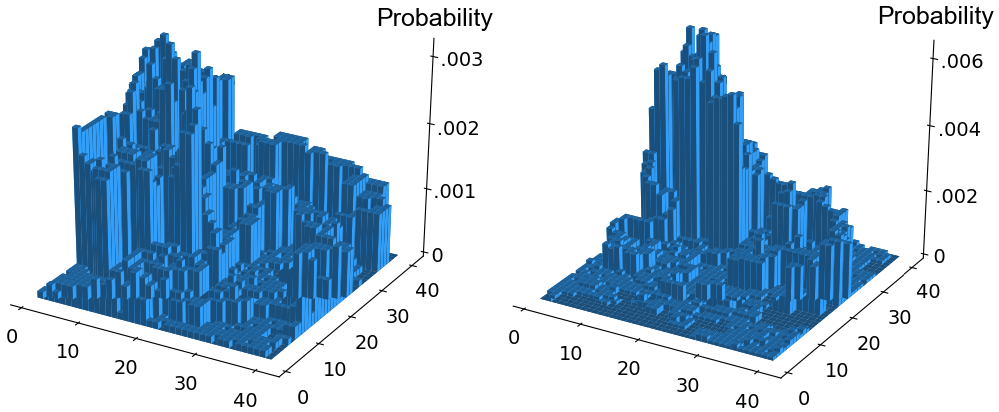}
\caption{The probability distribution P(x,y) for grid geometries representing perfect square mazes. Given enough unitary steps, the probability distribution in the system can hint at the location of F, and nearby paths leading away from F.}
\label{Fig:3D_PSM}
\end{figure}

As expected, the viability of the stable hybrid search completely deteriorates as the wall count approaches the maximum. In fact, for these higher wall counts, we found that the most common number of unitary steps to produce the fastest stable hybrid speeds was 0, which means that the fastest search is actually just a regular classical search, with no help from the quantum system. Now, to illustrate why the SQRW technique is failing to produce any sort of speedup, figure \ref{Fig:3D_PSM} shows some examples of probability distributions generated on perfect square mazes.

At first glance, these probability distributions seem to show some signs for a successful hybrid search. States closer to F have clearly higher probabilities, and even some paths are revealed in certain areas. In fact, these kinds of P(x,y) distributions $\textit{do}$ result in faster classical searches. That is to say, if we were given the P(x,y) distributions in figure \ref{Fig:3D_PSM} and used a measurement to determine the starting point for a classical search as per equation \ref{eq:steps}, we would get a speedup.

However, the number of unitary steps needed to reach probability distributions like those in figure \ref{Fig:3D_PSM} eliminate any hope of a successful hybrid algorithm. As it turns out, performing a SQRW on a perfect square maze geometry typically needs $\textit{more}$ than N$^2$/2 unitary steps before a viable P(x,y) is reached. Thus, in the amount of time it takes the quantum system to reach a desirable point, a classical search will have on average already found F. As a whole, the system never reaches a point where the cost in unitary steps is worth the tradeoff for a desirable P(x,y).

%
\section{Conclusions}
%

The results of this paper have showcased the effectiveness of the Scatter Quantum Random Walk scheme as a search algorithm on N$\times$N geometries. For open grids (no obstacles), it was shown that the quantum systems representing these geometries produce desirable probability distributions P(x,y), which when used in combination with classical searches, produce hybrid search algorithms which are faster than purely classical searches. In addition, it was shown that these hybrid searches are still viable even when F's location is random, and that a single set of governing rules can be used for blind hybrid search algorithms. This result is perhaps the most important, as it shows that using a SQRW as a searching algorithm is not limited to rigid geometries with strict restrictions on the locations where F may be located.

When random obstacles are introduced into these grid geometries as walls that remove connections between nodes, it was shown that a hybrid searching technique is still viable up to a certain number of walls. However, as the number of walls in the geometry approached the limit of the perfect square maze, all viability for a hybrid search algorithm broke down. Nevertheless, the success of these hybrid searches at low wall counts is promising, suggesting that geometries with minimal obstacles may be viable for a blind hybrid search algorithm as well.

\subsection{Understanding SQRW Resilience}
%

In order to understand the true viability of using these SQRWs in quantum algorithms, and the results from sections 3-6, we must take great care in noting their strengths and limitations. Specifically, recall how the 'effect' of the SQRW spreads throughout the geometry, traveling out radially one node per unitary step. The speed of this spreading effect is what drives the success of the SQRW, and is a critical feature that determines the viability of certain geometries: relative distance from F to all nodes.

Regardless of exact shape, the speed at which the effect of the SQRW can reach all of the states in the system largely determines whether or not a useful probability distribution is obtainable. This principle is reinforced by the results found in this paper as follows: 1) Nodes closer to the center of grids produce faster speeds 2) Lattice geometries produce faster speeds when compared to grids of similar total nodes 3) As obstacles are placed throughout a grid geometry, restricting the ways for the SQRW's effect to spread, average speeds slow down as a result.

At its core, the Scattering Quantum Random Walk scheme is simply a means of representing a geometry as a quantum system and manipulating probabilities. Whether the purpose for implementing a SQRW is a search algorithm, or perhaps a smaller component to a larger quantum algorithm, it is hopeful that these and other quantum random walks may find use in algorithms on near term quantum computers.

\section*{Acknowledgement}

We gratefully acknowledge support from the National Research Council Associateship Programs and funding from OSD ARAP QSEP program. Any opinions, findings, conclusions or recommendations expressed in this material are those of the author(s) and do not necessarily reflect the views of AFRL.

\end{document}